\documentstyle[preprint,aps]{revtex}

\begin{document}
\draft
\title{Energy distribution of charged dilaton black holes}
\author{S.~S.~Xulu\footnotemark[1] 
\footnotetext[1]{Dept of Applied Mathematics, University of Zululand,
P/Bag X1001, 3886 Kwa-Dlangezwa, SOUTH AFRICA 
~~E-mail: ssxulu@pan.uzulu.ac.za}}
\maketitle
 \begin{abstract}

 Chamorro and Virbhadra studied, using the energy-momentum complex of
 Einstein, the energy distribution associated with static spherically
 symmetric charged dilaton black holes for an arbitrary value of the
 coupling parameter $\gamma$ which controls the strength of the dilaton
 to the Maxwell field. We study the same in Tolman's prescription and
 get the same result as obtained by Chamorro and Virbhadra. The energy
 distribution of charged dilaton black holes depends on the value
 of $\gamma$ and the total energy is independent of this parameter.

 \end{abstract}
 \pacs{04.70.Bw,04.20.Cv}

The subject of energy-momentum localization is associated with some
degree of controversy. There are different opinions on this subject.
Contradicting the viewpoint of Misner et al.\cite{MTW73} that the energy is
localizable only for spherical systems, Cooperstock and Sarracino\cite{CS78}
argued that if the energy localization is meaningful for spherical systems
then it is meaningful for all systems.
Bondi\cite{Bond90} expressed that a non-localizable form of energy is 
inadmissible in relativity and its location can in principle be found.
Einstein energy-momentum complex was followed by   
many definitions of energy, momentum, and angular momentum proposed for 
a general relativistic system (see in \cite{ACV96} and references therein). 
There lies 
a dispute with the importance of nontensorial energy-momentum complexes whose
physical interpretation has been questioned by a number of physicists, 
including Weyl, Pauli and Eddington (see in \cite{ChaFe91}). There is a 
suspicion that, in a given spacetime, different energy distributions would be 
obtained from different energy-momentum complexes. Several examples of 
particular spacetimes (the Kerr-Newman, the Einstein-Rosen, and the Bonnor-Vaidya)
have been investigated and  different energy-momentum complexes are known to
give the same energy distribution for a given spacetime\cite{KSV}. 
Recently, Aguirregabiria et al.\cite{ACV96} showed that several energy-momentum complexes
coincide for any Kerr-Schild class metric. 
     
    Virbhadra and Parikh\cite{VP93}, using the energy-momentum complex of
Einstein, calculated the energy distribution with stringy charged black
holes and found that the entire energy is confined to interior of the 
holes. We\cite{xulu97} found the same result using the Tolman definition.
Chamorro and Virbhadra\cite{CV96}, using the energy-momentum complex of
Einstein, studied energy distribution for the Garfinkle-Horowitz-Strominger
(GHS) charged dilaton black holes\cite{GHS91}. In this paper we investigate the
same using Tolman energy-mometum complex to see whether or not one gets
the same result. This is the purpose of our present investigation.
We use the convention
that Latin indices take values from 0 to 3 and Greek indices values
from 1 to 3, and take $ G=1$ and $ c=1$ units.

The Garfinkle-Horowitz-Strominger\cite{GHS91} static spherically symmetric asymptotically
flat black hole is described by the line element
\begin{equation}
ds^{2} = (1 - \frac{r_+}{r })(1 - \frac{r_-}{r })^\sigma dt^2 
        - (1 - \frac{r_+}{r })^{-1}(1 - \frac{r_-}{r })^{-\sigma} dr^2 
        -  (1 - \frac{r_-}{r })^{1-\sigma} r^2 (d \theta^2 + sin^2 \theta d \phi^2)
\label{eqn1}
 \end{equation}
and the dilaton field $\Phi$ is given by
\begin{equation}
e^{2\Phi} = \left(1-\frac{r_-}{r}\right)^{\frac{1-\sigma}{\gamma}},
\label{eqn2}
\end{equation}
where
\begin{equation}
~~\sigma = \frac{1-{\gamma}^2}{1+{\gamma}^2}.
\label{eqn3}
\end{equation}
$r_-$ and $r_+$ are related to mass M and charge Q parameters as follows:
\begin{eqnarray}
M &=& \frac{r_+ + \sigma r_-}{2},\nonumber\\
Q^2  &=& \frac{r_+ r_-}{(1 + \gamma^2)}.
\label{eqn4}
\end{eqnarray}
Virbhadra\cite{Vir97} proved that for $Q=0$ the GHS solution yields the
Janis-Newman-Winicour solution\cite{JNW68} to the Einstein-Massless Scalar
equations. Virbhadra et al.\cite{Viretal97} showed that the Janis-Newman-Winicour
solution has a globally naked strong curvature singularity. However, 
the GHS solution ($Q \neq 0$) is a black hole solution.

Charged dilaton black holes have been a subject of study in many
recent investigations\cite{HW92}-\cite{SSAD}. A number of interesting properties 
of charged dilaton black holes critically depend on a dimensionless
parameter $\gamma$ which controls the coupling between the dilaton
and the Maxwell fields. The maximum charge, for a given mass, that 
can be carried by a charged dilaton depends on $\gamma$ \cite{HW92}.
When $\gamma \neq 0$, the surface $r=r_-$ is a curvature singularity
while at $\gamma = 0$ the surface $r=r_-$ is a nonsingular inner horizon
\cite{HH92}. Both the entropy and temperature of these black holes
depend on $\gamma$ \cite{HW92}. The gyromagnetic ratio for charged
slowly rotating dilaton black holes depends on parameter $\gamma$
\cite{HW92}. 
Chamorro and Virbhadra \cite{CV96}  showed, using Eintein's
prescription, that the energy distribution of charged dilaton black holes 
depends on the value of $\gamma$.

We start by transforming the line element $(\ref{eqn1})$  to 
quasi-Cartesian coordinates:
\begin{eqnarray}
ds^{2}  = (1 - \frac{r_+}{r }) (1 - \frac{r_-}{r })^\sigma dt^2
     &-&  (1 - \frac{r_-}{r })^{1-\sigma}    (dx^2 + dy^2 + dz^2)\nonumber\\
&-&\frac{(1 - \frac{r_+}{r })^{-1}(1 - \frac{r_-}{r })^{-\sigma}
        -(1 - \frac{r_-}{r })^{1-\sigma}}{r^2} (xdx + ydy + zdz)^2 ,
 \label{eqn5}
 \end{eqnarray}
according to                               
 \begin{equation}
r = \sqrt{x^2 + y^2 + z^2}, ~~ \theta = cos^{-1}\left(\frac{z}{\sqrt{x^2 + y^2 + z^2}}\right),    
~~ \phi = tan^{-1} (y/x) .
\label{eqn6}
 \end{equation}

Tolman's\cite{Tol34} energy-momentum complex is
\begin{equation}
{\cal{T}}_k{}^i = \frac{1}{8 \pi}U^{ij}_{k,j},
\label{eqn7}
\end{equation}
where
\begin{equation}
U^{ij}_{k} = \sqrt{-g} \left[-g^{pi}( -\Gamma^j_{kp}
               +\frac{1}{2} g^j_{k} \Gamma^a_{ap}
               +\frac{1}{2} g^j_{p} \Gamma^a_{ak})  
           + \frac{1}{2} g^i_{k} g^{pm}( -\Gamma^j_{pm}
               +\frac{1}{2} g^j_{p} \Gamma^a_{am}
               +\frac{1}{2} g^j_{m} \Gamma^a_{ap})\right],
\label{eqn8}
\end{equation}
 ${\cal{T}}_0^0$ is the energy density, ${\cal{T}}_0^{\alpha}$ are 
the components of energy current density,  ${\cal{T}}_{\alpha}^{0}$ are 
the momentum density components.  Therefore, energy $E$ for a stationary
metric is given by the expression
\begin{equation}
E   = \frac{1}{8\pi}\int\int\int U_{0,\ \alpha}^{0\alpha} \ dx dy dz
\label{eqn9}\\
\end{equation} 
After applying the Gauss theorem one has
\begin{equation}
E = \frac{1}{8\pi}\int\int U_{0}^{0\alpha} \mu_\alpha dS ,
\label{eqn10}\\
\end{equation} 
where $ \mu_\alpha = (x/r, y/r, z/r)$ are the three components 
of a normal vector over an infinitesimal surface element 
$ dS = r^2 sin \theta d\theta d\phi$.

The determinant of the metric tensor and its non-vanishing contravariant 
components are obtained by Chamorro and Virbhadra\cite{CV96}. 
To compute energy using Eq. $(\ref{eqn9})$
we require the following list of nonvanishing components  of the
Christoffel symbol of the second kind.
\begin{eqnarray}
\Gamma^1_{11} &=& x ( c_1 + c_2 x^2 ),
                        ~~~~~~~~~~~~\Gamma^2_{22} = y ( c_1 + c_2 y^2 ),
                        \nonumber\\
\Gamma^3_{33} &=& z ( c_1 + c_2 z^2 ),
                        ~~~~~~~~~~~~\Gamma^1_{22} = x ( c_3 + c_2 y^2 ),
                         \nonumber\\
\Gamma^2_{11} &=& y ( c_3 + c_2 x^2 ),
                       ~~~~~~~~~~~~\Gamma^1_{33} = x ( c_3 + c_2 z^2 ),
                        \nonumber\\
\Gamma^3_{11} &=& z ( c_3 + c_2 x^2 ),
                       ~~~~~~~~~~~~\Gamma^2_{33} = y ( c_3 + c_2 z^2 ),
                       \nonumber\\
\Gamma^3_{22} &=& z ( c_3 + c_2 y^2 ),
                        ~~~~~~~~~~~~\Gamma^1_{12} = y (c_4 + c_2 x^2 ),
                        \nonumber\\
\Gamma^1_{13} &=& z (c_4 + c_2 x^2 ),
                        ~~~~~~~~~~~~\Gamma^2_{21} = x (c_4 + c_2 y^2 ),
                        \nonumber\\
\Gamma^2_{23} &=& z (c_4 + c_2 y^2 ),
                        ~~~~~~~~~~~~\Gamma^3_{31} = x (c_4 + c_2 z^2 ),
                        \nonumber\\
\Gamma^3_{32} &=& y (c_4 + c_2 z^2 ),
                        ~~~~~~~~~~~~\Gamma^1_{00} = x c_5,
                        \nonumber\\
\Gamma^2_{00} &=& y c_5,
                       ~~~~~~~~~~~~~~~~~~~~~~~\Gamma^3_{00} = z c_5,      
                        \nonumber\\
\Gamma^0_{01} &=& x c_6,
                       ~~~~~~~~~~~~~~~~~~~~~~\Gamma^0_{02} = y c_6,
                        \nonumber\\
\Gamma^0_{03} &=& z c_6, \nonumber\\
\Gamma^1_{23} ~~&=& ~~~\Gamma^2_{13} ~~~= ~~~\Gamma^3_{12} ~~~=  ~~~c_2 x y z .
\label{eqn11}
\end{eqnarray}
where\\
\begin{eqnarray}
c_1 &=& \frac{1}{2 r^4 (r - r_-)} [2 r_+ r^2 +3r_- r^2 - 3 r_- r_+ r + 
                                     {r_-}^2 r_+ - {r_-}^2 r + 
                                     ( r_- r_+ - r_- r - r_+ r - r^2) 
                                      r_- \sigma] ,
                        \nonumber\\
c_2 &=& \frac{1}{2 r^4} \left[ \frac{2{r_+}^2r +6r_-r_+r - 3r_-r^2 - 
                                     3r_+r^2 - 3r_-{r_+}^2}{r - r_+}
                              + \frac{(2r_- -r_-\sigma)(r_-r_+ - r_-r - r_+r)}
                                     {r - r_-}
                         \right] ,
                        \nonumber\\
c_3 &=& \frac{1}{2r^4} [ r_-r + 2r_+r - r_-r_+ + (r - r_+)r_-\sigma ] ,
                        \nonumber\\
c_4 &=& \frac{1}{2r^2} \left[ \frac{r_- - r_-\sigma}{r - r_-} \right] ,
                        \nonumber\\
c_5 &=& \frac{(r -r_-)^{2\sigma - 1}}{2r^{2\sigma + 4}}(r - r_+) 
                        [ (r - r_-)r_+ + (r - r_+)r_-\sigma ] ,
                        \nonumber\\
c_6 &=& \frac{1}{2r^2} \left[ \frac{r_+}{r - r_+} + 
                            \frac{r_- \sigma}{r - r_-}
                       \right] .
\label{eqn12}
\end{eqnarray}
Using Eqs. $(\ref{eqn8})$ and $(\ref{eqn11})$ we obtain required components
 of  $U^{ij}_{k}$. These are  
\begin{eqnarray}
U^{01}_{0} = \frac{x}{r^4} \left[ r(\sigma r_- + r_+) - \sigma r_- r_+ \right],
 \nonumber\\
U^{02}_{0} = \frac{y}{r^4} \left[ r(\sigma r_- + r_+) - \sigma r_- r_+ \right],
\label{eqn13}\\
U^{03}_{0} = \frac{z}{r^4} \left[ r(\sigma r_- + r_+) - \sigma r_- r_+ \right].   
\nonumber
\end{eqnarray}
Now using  $(\ref{eqn13})$ with $(\ref{eqn4})$  in $(\ref{eqn10})$ we get
\begin{equation}
E(r) =  M - \frac{Q^2}{2r} (1-\gamma^2) .
\label{eqn14}
\end{equation}
Thus, we get the same result as Chamorro and Virbhadra\cite{CV96} obtained 
using  the Einstein energy-momentum complex. This is against the ``folklore''
that different energy-momentum complexes could give different and hence 
unacceptable energy distribution in a given spacetime. For the 
Reissner-Nordstr\"{o}m metric one gets $E = M -Q^2/2r$, which is the same
as obtained by using several other energy-momentum complexes\cite{ACV96}
and definitions of Penrose as well as Hayward\cite{TodHay}. $E(r)$, given
by $(\ref{eqn14})$, can be interpreted as the ``effective gravitational mass''
that a neutral test particle ``feels'' in the GHS spacetime. The ``effective
gravitational mass'' becomes negative at radial distances less than
$Q^2 (1-\gamma^2)/2M$.

\acknowledgments

Thanks are due to T. A. Dube and K. S. Virbhadra for their guidance.

\end{document}